\documentstyle[preprint,eqsecnum,aps]{revtex}

\begin{document}
\draft
\title{Calculations on Electronic States in QDs with Saturated Shapes}
\author{Wei Cheng}
\address{Key laboratory in University for Radiation Beam Technology and Materials\\
Modification,Institute of Low Energy Nuclear Physics, \\
Beijing Normal University, Beijing 100875, P. R. China}
\author{Shang-Fen Ren}
\address{Department of Physics, Illinois State University, Normal,\\
Illinois 61790-4560}
\date{\today}
\maketitle

\begin{abstract}
Electronic States of Si and Ge QDs of 5 to 3127 atoms with saturated shapes
in a size range of 0.57 to 4.92 nm for Si and 0.60 to 5.13 nm for Ge are
calculated by using an empirical tight binding model combined with the
irreducible representations of the group theory. The results are compared
with those of Si and Ge quantum dots with spherical shape. The effects of
the shapes on electronic states in QDs are discussed.
\end{abstract}


\preprint{HEP/123-qed}

\section{Introduction}

Semiconductor quantum dots (QDs) have attracted much research attention in
recent years because of their importance in the fundamental understanding of
physics and potential applications \cite{Yoffe}. One of the most important
properties of semiconductor QDs is the change of the electronic band
structure of QDs when the size of the QDs changes\cite
{wang,brus,ali,SYRSSC97,SYRPRB97}. The blue shifts of the bandgap of
semiconductor QDs has been demonstrated by numerous experimental
observations \cite{Yoffe}, and there exist many successful theoretical
investigations on this and other related important physical properties.
Among theoretical models, the effective mass approach (EMA) which predicted
the increase of the band gap as the size of the QDs decreases\cite
{wang,brus,ali}, which is simple to understand and provides qualitatively
correct description of the increase of the band gaps. The electronic
structures of semiconductor QDs has also been investigated by microscopic
models \cite
{wang,brus,ali,SYRSSC97,SYRPRB97,SYRJAP97,SYRPRB92,kumar,rama,pan,bahel},
including an empirical tight-binding approach combined with the irreducible
representations of the group theory\cite{SYRPRB92} that has obtained many
interesting results. One of the most important results is the existence of a
critical size in spherical semiconductor QDs\cite{SYRJAP97}. In above
models, the shape of the QDs treated is taken as spherical. More shapes and
structure of Si and Ge QDs are proposed\cite{Yoffe,ho}, i.e., a complicated
structure for Si$_{10}$, and between Si$_{20}$ and Si$_{30}$ alters from
being elongated to spherical\cite{parent}. In this work, we have calculated
the electronic states of QDs with saturated shapes at different sizes. Our
results are compared with those of the spherical shape, and the effects of
the shapes of QDs on electronic states in QDs are discussed.

This paper is organized as the following: first, we will describe the
saturated shape of the QDs and briefly describe our theoretical approach,
then we will show our results and have discussions.

\section{Saturated QDs and theoretical approach}

The semiconductor materials we discuss in the work, Si and Ge, have
diamond structures. In QDs, this structure remains. The saturated shapes
of semiconductor QDs discussed in this work are built up in the following
way: first we start from a center atom with its four nearest neighboring
atoms. This is a saturated QDs with the minimum size. Then we add all the
next neighbors to the four surface atoms to form the next saturated QD,
which has 17 atoms in total. Then we add all the next neighbors to the
twelve surface atoms to form the next saturated QD again, which now has 41
atoms. The larger saturated QDs are built up by repeating this procedure.
The number of Si (Ge) atoms we calculated here are 5, 17, 41, 83, 147, 239,
363, 525, 729, 981, 1285, 1647, 2071, 2563, and 3127. In Fig. 1, we show the
shapes of the saturated QDs with 363, 1647, and 3127 atoms. The shape is a
truncated cube. For large QD the four small and four large triangles will
have nearly the same size, and six rectangles will be six squares. There are
a few features of the saturated shape of QDs that we want to mention. First,
when the number of atoms in QDs is equal or less than 17, the structure of
the saturated QDs is exactly the same as the spherical QDs. So our
calculated results should agree with the existing results of spherical QDs.
At the beginning of this work, we did check it, and they agree exactly.
Second, the saturated shape is not spherical, so the distances from surface
atoms to the center of the QDs are not the same. We have defined the radius $%
r$ of the QDs in the following way: $Nm=\rho (\frac 43\pi r^3)$, when $N$ is
the total number of atoms in the QD, $m$ is the mass of a $^{70}$Ge atom, $%
\rho $ is the density of the bulk material, and $r$ is an equivalent radius
of the spherical QDs with the same number of atoms. Third, one important
feature of the saturated shape is that even though it looks like more
complicated, it keeps the same $T_d$ symmetry of the bulk material. Because
of this, the irreducible representations of the group theory applied in the
spherical QDs can also be applied in QDs with saturated shape. This makes
the comparison with the results more convenient.

We employed the same empirical tight-binding approach and parameters as the
calculations on spherical QDs \cite{SYRSSC97,SYRPRB97,SYRJAP97,SYRPRB92}.
This empirical tight-binding model reproduces the correct bandgap of bulk Si
and Ge in the limit of infinite clusters by construction\cite{vogl}, and its
simplicity makes the calculation for very large QDs feasible. We also made
the following assumptions following the calculations of spherical QDs:
first, we take the hydrogen saturated approximation, i.e., the dangling
bonds of Si and Ge atoms at the surface of the QDs are terminated with
hydrogen atoms; second, atoms in QDs take the diamond lattice sites. The
hydrogen saturated dangling bonds at the surface of the QDs are assumed to
have the same length as the nature H-Si or H-Ge bond length ($%
d_{H-Si}=0.148nm$, and $d_{H-Ge}=0.153nm$). Then the electronic structures
are evaluated by using the empirical Hamiltonian\cite{vogl}, that produces
the accurate valence bands and good conduction bands near the fundamental
band gap for bulk Si and Ge. We have considered five basis orbitals per Si
or Ge atom for the Hamiltonian: $s$, $p_x$, $p_y$, $p_z$, and an excited $%
s^{*}$ state. In this Hamiltonian, only on-site and nearest neighbor
interaction matrix elements are considered as non-zero. Each hydrogen atom
has only one single $s$ orbital. Since the hydrogen free atom energy level
(-13.6 eV) is close to the $s$-state energy level of Si ($-$13.55 eV), the
on-site $s$ energy level of hydrogen is taken to be the same as that of Si.
The nearest neighbor matrix elements $V_{H-Si}$ ($V_{H-Ge}$ between H and Si
(Ge) are taken to be the same as Si-Si (Ge-Ge), but scaled inversely as the
square of the bond length $d$ according to Harrison's rule \cite{Harrison}.
The QDs we have calculated ranging from five Si (Ge) atoms with twelve
surface hydrogens to 3127 Si (Ge) atoms with 1188 hydrogens. Without group
theory the dimension of the largest Hamiltonian matrix for the saturated QD
of 3127 atoms is 16823=3127$\times $5+1188. Such large matrices are
difficult to be diagonalized directly, so the projection operators of the
irreducible representations of the group theory \cite
{SYRSSC97,SYRPRB97,SYRJAP97,SYRPRB92} are employed to reduce the
computational intensity. By employing the group theory, for example, the
above matrix of size of 16823 can be reduced to five matrices in five
different representations of A$_1$, A$_2$, E, T$_1$, and T$_2$, with the
sizes of 849, 568, 1397, 1962, and 2242 respectively. Therefore, the
original problem is reduced to a problem that can be easily handled by most
reasonable computers. Furthermore, the employment of the group theory proves
to have played a much more important role than expected. Not only it allows
the investigation of electronic states in QDs with a much larger size, but
also it allows the investigation of electronic states in QDs with different
symmetries. This group theory formalism has been also used in calculations
of phonon modes in semiconductor QDs. These investigations lead to many
interesting physics that otherwise can not be revealed\cite
{SYRSSC97,SYRPRB97,SYRJAP97,SYRPRB92,QD1,QD2,QD3,QD4,QD5}.

\section{Results and Discussion}

With this model, we have calculated the electronic states in saturated Si
and Ge semiconductor QDs. Our results for Si saturated QDs are plotted in
Fig. 2 (a) and (b) for Si. Fig. 2 (a) shows the calculated lowest unoccupied
energy levels for saturated Si QDs ranging from 5 to 3127 Si atoms. Two
levels are shown for each of the five different irreducible representations.
Fig. 2 (b) shows the calculated highest occupied energy levels for the same
sets of Si QDs, and also two levels are shown for each of the five different
irreducible representations. 


Our results show that when the QDs have only 5 or 17 Si (Ge) atoms, the
results are exactly the same as those of spherical QDs (the corresponding
results for Si spherical QDs are in Fig. 2 of reference \cite{SYRPRB97}, and
those of Ge spherical QDs are in Fig. 2 of reference \cite{SYRSSC97}). This
is in this size range the saturated QDs and the spherical QDs have exactly
the same structures. When the size of QDs increases, the shape of saturated
QDs are different from the spherical QDs, so the electronic structures of
the saturated QDs have obvious differences from those of spherical QDs.

\subsection{Lowest unoccupied states of Si saturated QDs}

From Fig. 2 (a) we see that all of the lowest unoccupied levels go up
monotonically as the QDs decreases, while the very lowest three are always
one from A$_1$ E, and T$_2$ each for QDs larger than 2.0 nm in diameter.
These levels are well separated from all other energy levels above them but
very close to each other. This is the same as the spherical QDs \cite
{SYRPRB97}, and can be explained as that these three states are directly
developed from the conduction minimum in the bulk. When the QDs size is big,
the coupling between different conduction minimum can be neglected, and all
these three lowest unoccupied states have almost the same energy. As the
size of QDs decrease, the coupling between different states increases, the
originally almost indistinguishable energy levels of A$_1$, E, and T$_2$
develop to three separated ones. The other character similar to spherical
QDs is as the QD size greater than 2.0 nm the two lowest T$_1$ modes have
nearly the same energy.

\subsection{Highest occupied states of Si saturated QDs}

Fig. 2 (b) shows the highest occupied levels of above Si saturated QDs. We
see that all the occupied levels go down monotonically as the size of the
QDs decreases. On this figure, the highest occupied level is always a T$_2$
level, and the next one is always a T$_1$ level. Different from those of
spherical Si QDs (Fig. 2 (b) of Reference \cite{SYRPRB97}) where there are
two crossovers of the T$_1$ and T$_2$ states, the first is the highest
occupied state changes from a T$_2$ state to a T$_1$ state in the size range
between 1.08 and 1.41nm, then the highest occupied state changes from a T$_1$
state to a T$_2$ state in the size range of 2.03 to 4.91nm. In our figure,
the highest occupied state is always a T$_2$ state, and there is no
crossover of T$_2$ and T$_1$ states in the size range we calculated here.
This is because in the saturated QDs there are more hydrogen atoms, the
interaction of hydrogen can increase the T$_2$ energy levels.

\subsection{Lowest unoccupied states of Ge saturated QDs}

Fig. 3 (a) shows the lowest unoccupied levels of Ge saturated QDs that go up
monotonically as the QDs decreases. The very lowest two are always from A$_1$
and T$_2$ each for QDs larger than 1.5 nm in diameter. These levels are well
separated from all other energy levels above them but very close to each
other. This is the same as the spherical QDs (Fig. 2 (a) of Reference \cite
{SYRSSC97}), and can be explained as that these two states are directly
developed from the conduction minimum in the bulk, which is at L point in
the Brillouin Zone. When the QDs size is big, the coupling between different
conduction minimum can be neglected, and these two lowest unoccupied states
have almost the same energy. As the size of QDs decrease, the coupling
between different states increases, the originally almost indistinguishable
energy levels of A$_1$ and T$_2$ develop to two separated ones. The other
character similar to spherical QDs is as the QD size greater than 3.5 nm the
two lowest T$_1$ modes have nearly the same energy.

\subsection{Highest occupied states of Ge saturated QDs}

Fig. 3 (b) shows the lowest unoccupied levels of above Ge saturated QDs. All
the occupied levels go down monotonically as the QDs decreases. On this
figure, the highest occupied level is a T$_1$ level when the size of QDs is
large, and the next one is a T$_2$ level. On the other hand, when the size
of QDs is small, the highest unoccupied level is a T$_2$ level, and the next
one is a T$_1$ level. We see obviously that there is a crossover of the T$_1$
and T$_2$ states at the size of 3 nm in diameter. This is different from
those of spherical Ge QDs (Fig. 3 (b) of Reference \cite{SYRSSC97}), where
the crossover is at 2 nm and the T$_1$ level is always a highest one in the
same size range. Since we know that in bulk material, the highest unoccupied
level is a T$_2$ level, we can image that there is another crossover of T$_1$
and T$_2$ at a larger QD size.

\subsection{More about the Crossover of Highest Occupied T$_2$ and T$_1$
States}

We see from above that the highest occupied levels for both Si and Ge
saturated QDs are different from those of corresponding spherical QDs. The
symmetry of the highest occupied levels is important, because there could
exist a critical size in these semiconductor QDs that when the size of QDs
decreases pass the size, the originally direct semiconductor becomes
indirect and the originally indirect semiconductor becomes less indirect 
\cite{SYRJAP97}. Our results show that when the shape of QDs is different,
the crossover of the T$_2$ and T$_1$, if they exist, will happen at
different size range. Therefore, if the crossover of the T$_2$ and T$_1$
states is desired, the selection of the shapes might help.

\section{Summary}

In summary, we have calculated electronic states in Si and Ge QDs of 5 to
3127 atoms with saturated shape in a size range of 0.6nm to 51.33nm in
diameter. The calculated results are compared with those of corresponding
QDs in spherical shape, and similarities and differences are discussed in
detail. Our results show that the influence of the shape of QDs on the
electronic states is important, and it may play an important role in the
band gap property of semiconductor QDs.

\acknowledgements

This research is supported by the National Science Foundation 
(INT0001313), and WC is also supported by 
the National Natural Science Foundation of China under Grant No. 10075008,
Visiting Scholar Foundation of Key Lab in University, Research Fund for the
Doctoral Program of Higher Education under Grant No. 20010027005, and
Excellent Young Teacher Foundation of the Education Ministry of China. We
thank Professor Shang-Yuan Ren for helpful discussions.

\newpage

\begin{figure}[tbp]
\caption{Schematic illustration of saturated shape which is a truncated cube
with six rectangles, four small, and four large triangles (a). Surface atoms
in saturated QD viewed from one side of the cube for 363 (b), 1647 (c) and
3127 (d) atom QDs. (e) and (f) are the 3127 atom QD viewed in two different
dirrections. It can be shown that for large QD the four small and four large
triangles will have nearly the same size, and six rectangles will be six
squares.}
\label{fig1}
\end{figure}

\begin{figure}[tbp]
\caption{(a) The two lowest unccupied energy levels for each of the five
irreducible representations as functions of the saturated Si QDs size. Note
that the A$_1$, E, and T$_2$ lowest unoccupied states are the very lowest
three states and are almost well spperated from other levels above. (b) The
two highest occupied energy levels for each one of the five irreduciple
representations as functions of the size of above saturated Si QDs. Note
that the T$_2$ and T$_1$ higest occupied states are almost always teh very
highest two levels and well seperated from other levels below.}
\label{fig2}
\end{figure}

\begin{figure}[tbp]
\caption{(a) The two lowest unccupied energy levels for each of the five
irreducible representations as functions of the saturated Ge QDs size. Note
that the A$_1$ and T$_2$ lowest unoccupied states are the very lowest two
states and are almost well spperated from other levels above. (b) The two
highest occupied energy levels for each one of the five irreduciple
representations as functions of the size of above saturated Ge QDs. Note
that the T$_2$ and T$_1$ higest occupied states are always the very highest
two levels and well seperated from other levels below, and there is a
crossover of the T$_2$ and T$_1$ states at the size of about 2 nm in
diameter.}
\label{fig3}
\end{figure}

\newpage

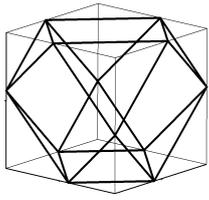
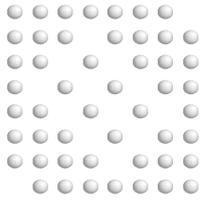
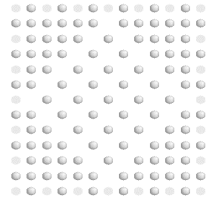
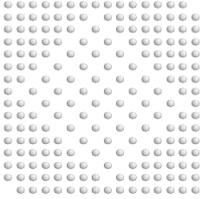
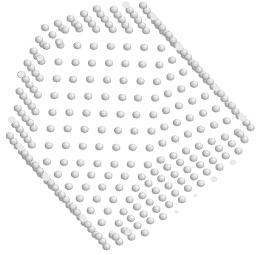
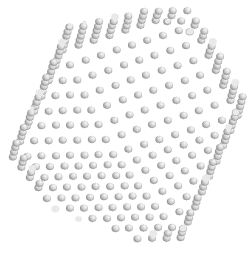
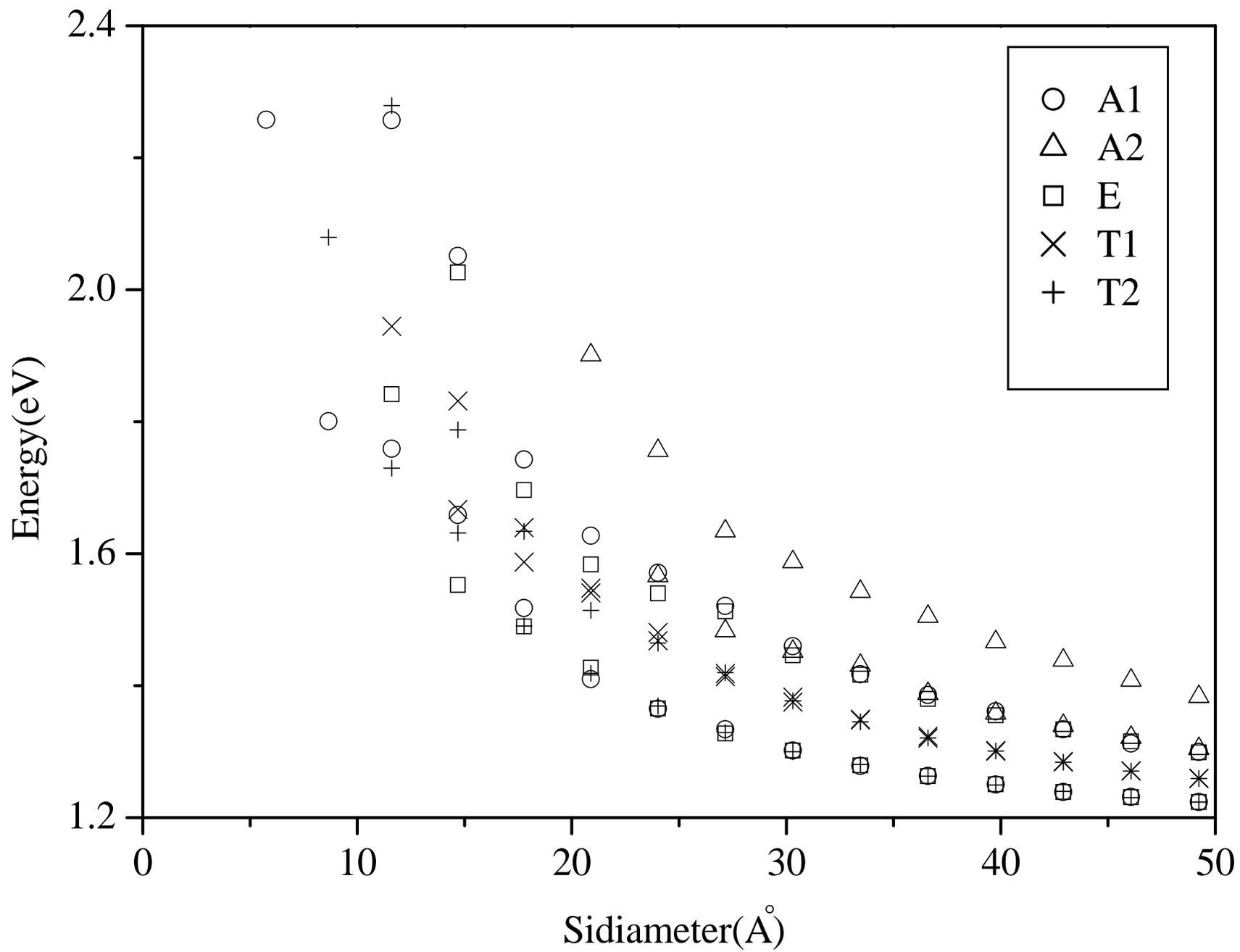
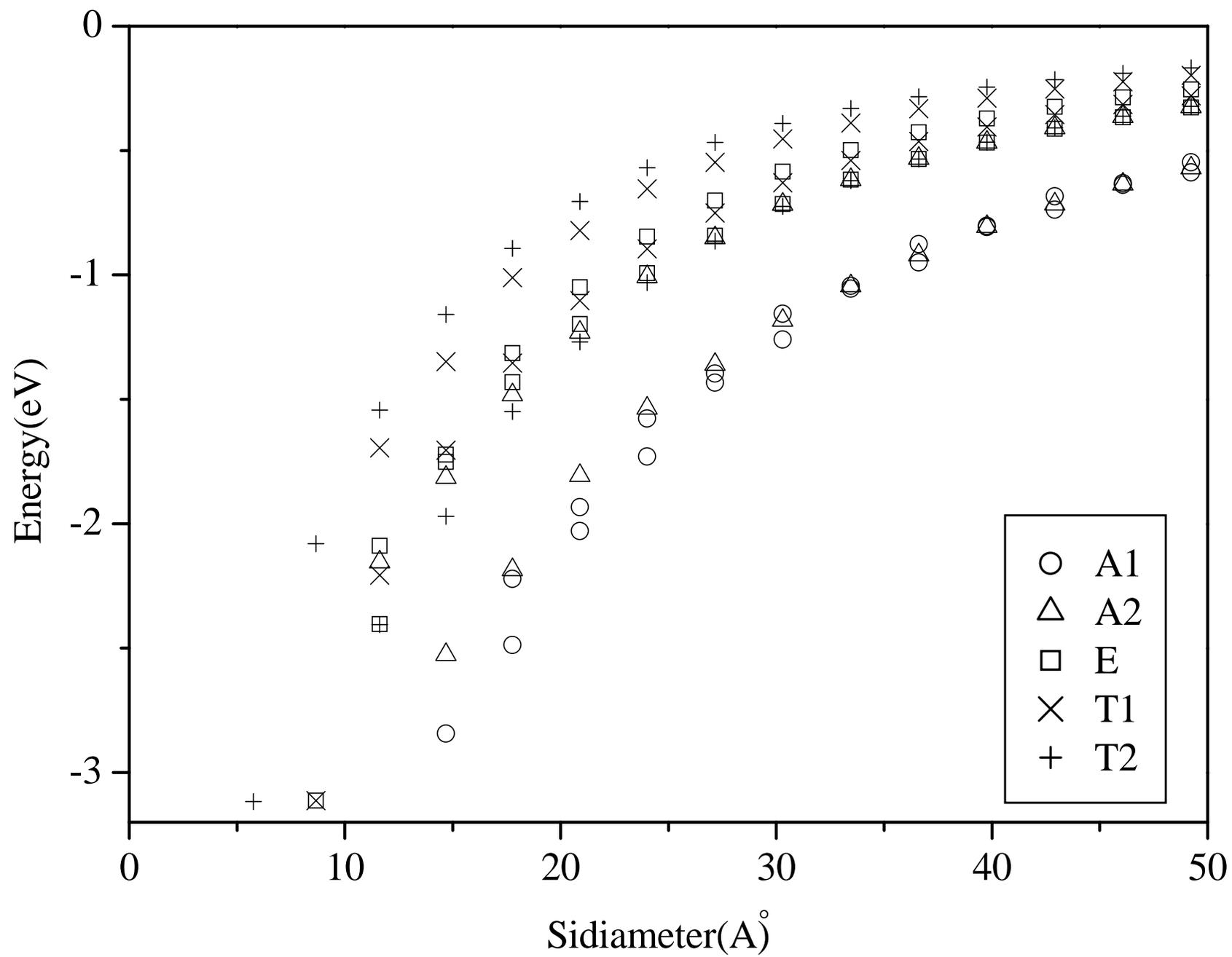
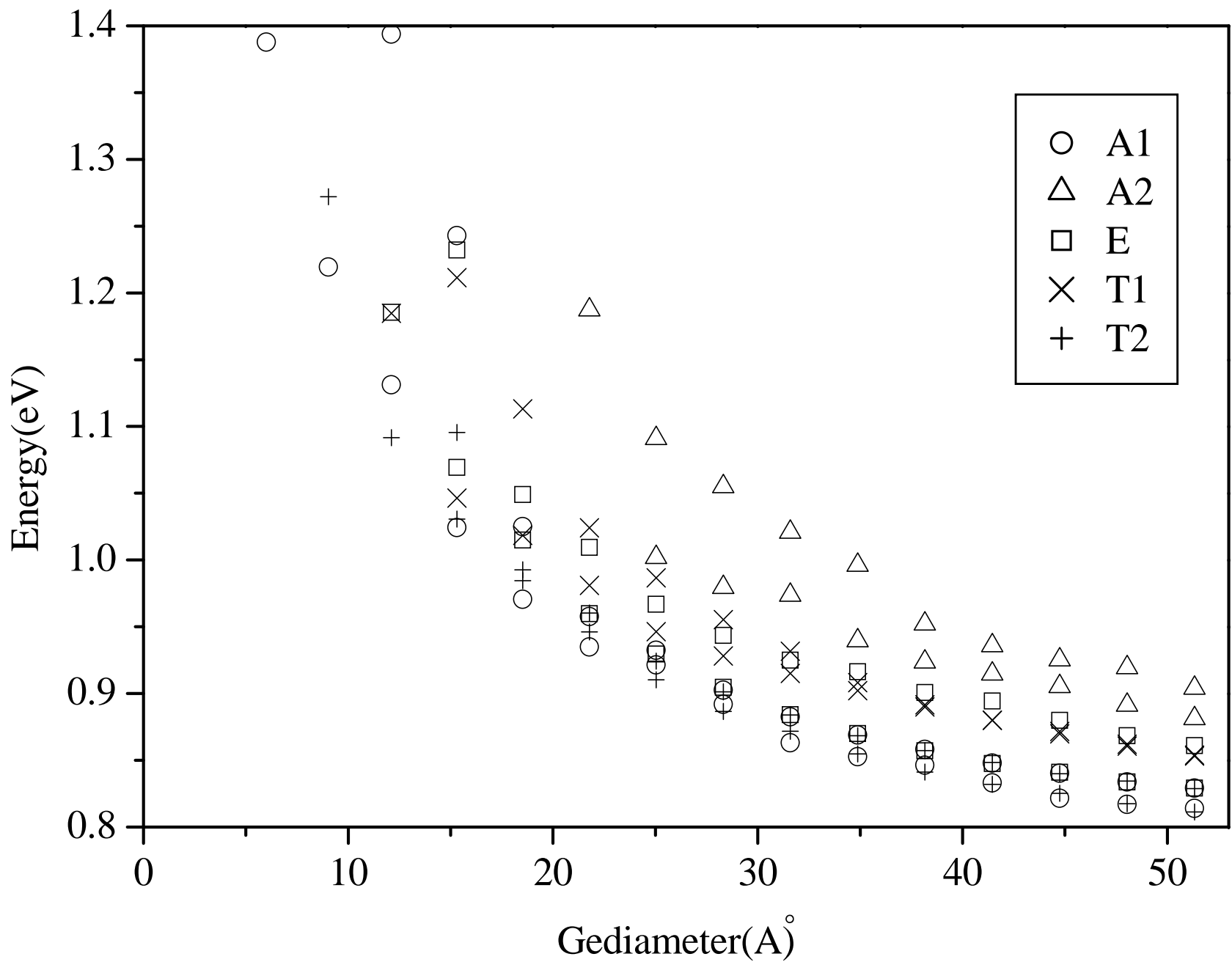
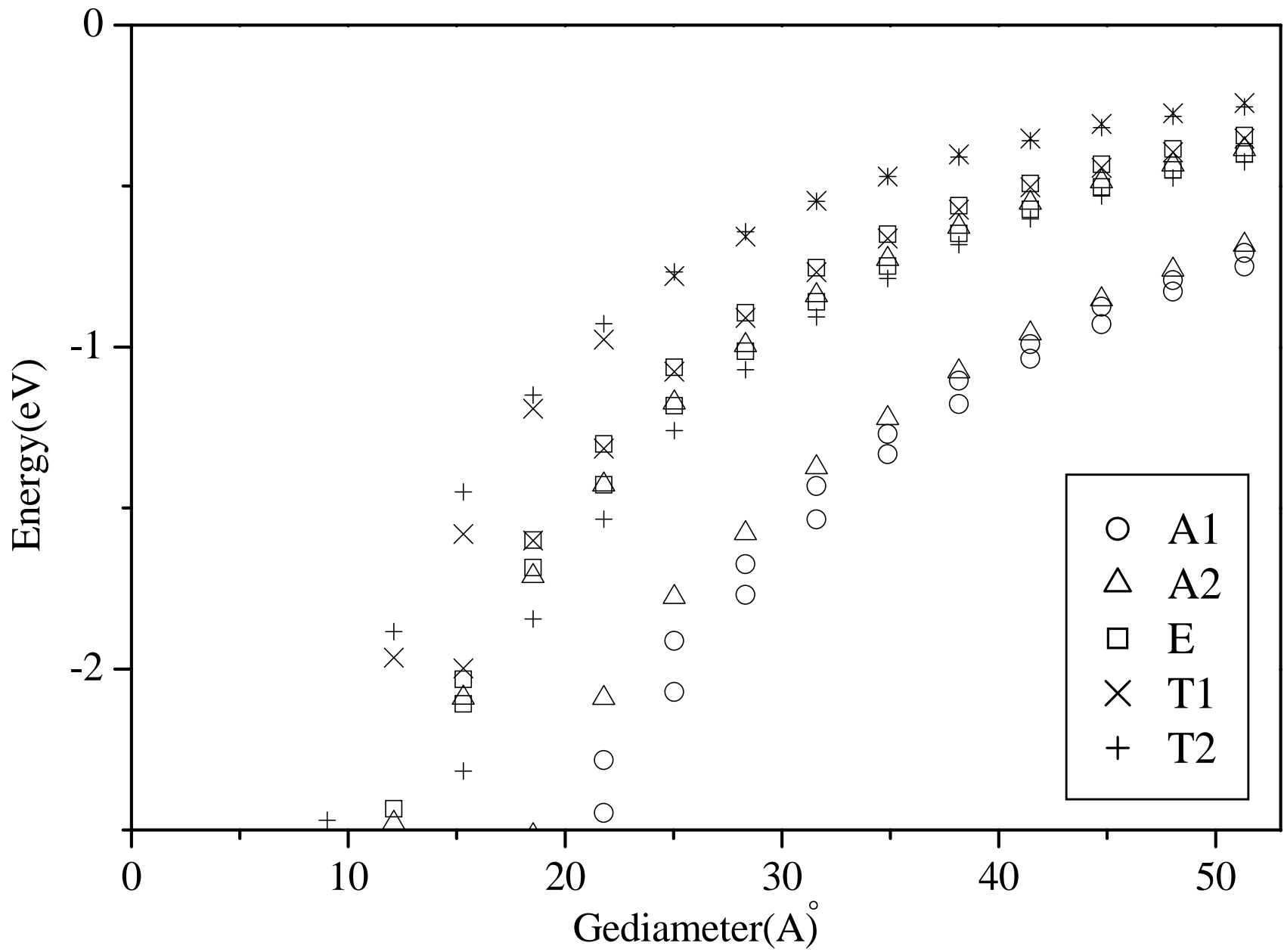
\end{document}